\documentclass[twocolumn,apjl]{emulateapj} 
\usepackage{times}
\usepackage{natbib}

\journalinfo{The Astrophysical Journal, 699:L161-L164, 2009, July 10}
\submitted{Received 2009 March 31; accepted 2009 May 29}

\newcommand{\oiii}{{[O\,{\sc iii}]}}
\newcommand{\oii}{{[O\,{\sc ii}]}}
\newcommand{\nii}{{[N\,{\sc ii}]}}
\newcommand{\hb}{{H$\beta$}}
\newcommand{\ha}{{H$\alpha$}}
\newcommand{\oh}{12\,+\,log(O/H)}
\newcommand{\ohe}{\mbox{12\,+\,log(O/H)\,=\,}}
\newcommand{\lam}{$\,\lambda$}
\newcommand{\llam}{$\,\lambda\lambda$}
                                                                                
\begin{document}
\title{First direct metallicity measurement of a lensed star-forming galaxy at $z=1.7$\footnotemark[*]}
\footnotetext[*]{Based on data collected 
at {\it Subaru} Telescope, which is operated by the National Astronomical Observatory of Japan.}

\author{T.-T. Yuan and L. J. Kewley}

\affil{Institute for Astronomy, University of Hawaii, 2680 Woodlawn
Drive, Honolulu, HI 96822, USA}

\begin{abstract}
We present the rest-frame optical spectrum of a strongly lensed galaxy at redshift $z =1.7$ behind the cluster Abell 1689.
We detect  the temperature sensitive auroral line \oiii\lam4363, 
which allows the first {\it direct} metallicity measurement for galaxies at $z >1$.
Our high signal-to-noise spectrum indicates that the target is an extremely low metallicity star-forming galaxy, 
with \ohe $7.5^{+0.1}_{-0.2}$ from the direct  $T_{e}$-based method.
We estimate an intrinsic  absolute $B$-band magnitude of $M_{B}=-18.3 \pm 0.1$, 
with a stellar mass of 4.4$\pm$1.2$\times$10$^{8}$ M$_{\odot}$.  This galaxy extends the 
luminosity---metallicity relation of star-forming galaxies at $z>2$ by more than an order of magnitude.
Given the double-nuclei like morphology and the velocity profile of \ha, 
we tentatively suggest that it could be a merger or a proto-rotating disk galaxy.  
\end{abstract}

\keywords{galaxies: abundances --- galaxies: evolution --- galaxies: high-redshift}

\section{Introduction}
Five years ago,  our knowledge about galaxy evolution still had a glaring gap between redshifts 1.5 and 3, termed the redshift desert. The desert is even more desolate for galaxy chemical evolution studies, where high signal-to-noise (S/N) spectra are needed for reliable measurements.  However, great 
progress has been made in the last few years in studying galaxies at z$\sim$ 2
and we are beginning to harvest the ``desert".  To date, metallicities are available for 
a few tens of  individual galaxies  at $1.5<z<3$ \citep{Shapley04, Erb03, Erb06, Hayashi09}. 
These galaxies are selected using broad-band colors either in the UV  \citep[Lyman Break technique;][]{Steidel96,Steidel03}
 or using $B$, $z$, and $K$-band colors \citep[BzK selection;][]{Daddi04}.  
 The Lyman break and $BzK$ selection techniques favor the galaxies that are luminous in the UV or  blue and could therefore be biased against low  luminosity (low-metallicity) galaxies, and dusty (potentially metal-rich) galaxies.

To date, metallicity measurements for high-redshift ($z \geq 1$) galaxies have been made 
via indirect methods from the ratios of strong nebular lines.  In local galaxies, the preferred method for determining metallicity
is via the ``direct" $T_{e}$ method which relies on the electron temperature sensitivity of the \oiii\lam4363 line.
At high redshift,  metallicity measurement via the direct method has not been possible because the 
\oiii\lam4363 line used to derive the electron temperature is very weak. 
The highest redshift that the \oiii\lam4363 line has been reported is at $z \sim 0.8$ by \citet{Hoyos05} and \citet{Kakazu07}.  

Strong lensing by galaxy clusters offers an alternative tool to explore the metallicity of galaxies at high-redshift.
One advantage of lensing selected galaxies is that they are immune to color and luminosity biases.  
Previously, only five lensed galaxies had been observed with sufficient spectral coverage and S/N to yield metallicities
 \citep{Yee96,Pettini00,Pettini02,LB03,Nesvadba07,Stark08}. Compared with Lyman break selected 
galaxies in the same redshift and luminosity range, the five lensed objects 
span a substantially broader metallicity range (\oh $\sim$7.1---9.1), indicating that current metallicity history 
studies at $z > 1.5$  (\oh $\sim$8.4---8.8) may not be sampling the full metallicity range at this redshift. 
It is worth noting that the metallicity range quoted for the lensed galaxies is extremely uncertain, depending
on various assumptions about reddening and metallicity calibrators.
Another great advantage of gravitational lensing is its natural magnification of the source.
The flux of the lensed background galaxy is commonly
 boosted by a factor of 1-3 mag or more around the critical lines of the cluster center.
Optical spectroscopic surveys of high-z galaxies near cluster critical lines have already yielded a large catalog of lensed galaxies beyond z $\sim$1
 \citep[e.g.,][]{Broadhurst05, Richard08, Frye07, Limousin07}. Unfortunately,  
spectra from these surveys cannot be directly utilized in metallicity studies because the commonly used metallicity diagnostic emission lines have shifted into 
the near infrared (near-IR) at  $z>1$. 

With the revolutionary multi-object cryogenic near-IR spectrographs that have been recently installed on 8-10m class telescopes,  
high efficiency near-IR spectroscopy  is now available.   We carried out 
the first near-IR spectroscopic survey of lensed galaxies behind lensing clusters, aimed at measuring the metallicity of galaxies
at $1.5<z<3$.  This Letter reports the rest-frame optical spectrum 
of a strongly lensed star-forming galaxy at $z = 1.7$ behind the Abell cluster 1689 from our ongoing survey.  
The galaxy cluster Abell 1689 was chosen as our first target because it has the largest Einstein radius
of $\sim 50\,\!\!^{\prime\prime}$ \citep{Broadhurst05}, 
and nearly $\sim$ 100 spectroscopically confirmed lensed images \citep{Limousin07}.
The galaxy in question is designated as ID 22.3 in the catalog of \citet{Broadhurst05}, 
we refer to it as Lens22.3 hereafter.  Throughout we use the standard lambda cold dark matter ($\Lambda$CDM) cosmology with $H_0=70$ km s$^{-1}$ Mpc$^{-1}$,
$\Omega_M=0.3$ and $\Omega_{\Lambda}=0.7$.

\section{Observations and Data Reduction}
%%%
\begin{figure}[!ht]
\epsscale{1.1}
\plotone{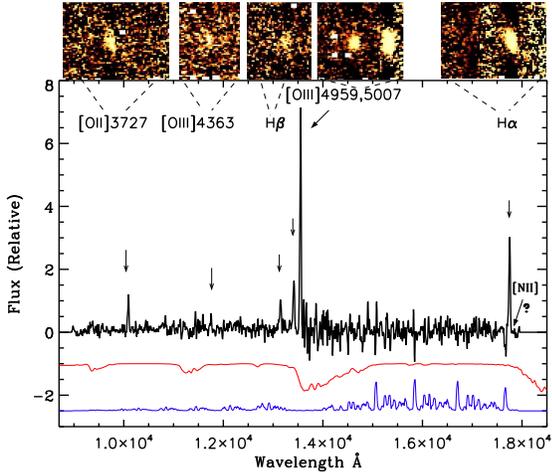}
\caption{Observed {\it Subaru} MOIRCS spectrum of Lens22.3 at z= 1.7.  
The extracted one-dimensinoal spectrum is shown in black, with the identified emission lines labeled.  The sky telluric absorption and OH emission
lines are shown in red and blue, respectively.  The two-dimensional spectrum for each of the detected lines is shown on top.}
\label{fig:spec}
\end{figure}
 %%%
 %%%%
%%% 
The data were obtained on 2008 March 26th,  using the 
Multi-Object InfraRed Camera and Spectrograph (MOIRCS) at {\it Subaru} \citep{Ichikawa06}, 
in the spectroscopic mode \citep{Tokoku06}.    We used the  low-resolution grism zJ 500 with a nominal 
resolution of 500 and  a wavelength coverage of 0.9---1.78 $\mu$m.
We designed two masks targeting  the spectroscopically confirmed lensed images in the field of Abell 1689.
We  obtained useable data for one mask. 
During the observation for this mask,  the seeing varied between 0.$\!\!^{\prime\prime}$3 and 0.$\!\!^{\prime\prime}$6,  and the airmass 
 ranged between 1.1 and 1.7. Lens22.3 was placed at one of the center slits of the mask, with a slit width of $\sim 1\, \!\!^{\prime\prime}$ and a slit length of 
 $\sim 9\, \!\!^{\prime\prime}$.  
The telescope was nodded along the slit in a ABAB mode with a dithering length of 2.$\!\!^{\prime\prime}$5. 
The exposure time for each individual frame was 400-600 seconds, resulting in a total integration time of 6800 seconds. 

We reduced the data using the standard {\sc iraf} and IDL routines. 
Sky OH lines were removed by subtracting the closest dithering pair A and B. 
A scaling factor between each pair was determined to optimize the sky subtraction. 
The sky subtracted A-B pair was flat-fielded, cleaned for comic rays, and then traced and rectified using a dome flat spectrum.  The final two-dimensional spectrum was created by co-adding the reduced individual pairs. 
 Wavelength calibration was carried out using the strong OH lines between 0.9 and 1.8 $\mu$m.
A F5V standard star was observed at approximately the same airmass as the target. The Telluric absorption
was obtained by dividing the star spectrum with a fitted blackbody curve. The resulting stellar spectrum
was then divided into the object spectrum to correct for telluric absorption.

\section{Analysis and Results}
\subsection{Metallicity}

\begin{deluxetable}{lccccc}
\tabletypesize{\footnotesize}
\tablewidth{0pt}
\tablenum{1}
\tablecolumns{6}
\tablecaption{Observed Line Intensities and Ratios\label{tab:obs}}
\tablehead{
\colhead{Ion Line} & 
\colhead{$z_{obs}$}&
\colhead{Flux} &
\colhead{Flux (corrected)} &
\colhead{1$\sigma$} &
\colhead{S/N} \\
\cline{1-6}\\
\colhead{(1)} &
\colhead{(2)} &
\colhead{(3)} &
\colhead{(4)} &
\colhead{(5)} &
\colhead{(6)} \\
}
\startdata
\nii\lam6584& \nodata&   $<$0.009 &$<$0.025 &\nodata & \nodata \\ 
\ha\lam6563 & 1.705  &3.12 &5.03&$\pm$0.4 & $13$  \\ 
\oiii\lam5007& 1.706 & 6.45 &6.45& $\pm$ 0.3 & $15$  \\ 
\oiii\lam4959 &1.705 &  1.98 &1.98& $\pm$0.3 & $10$  \\ 
\hb\lam4861 &1.705   &1.0 &1.0& $\pm$0.1 & $5$  \\ 
\oiii\lam4363 & 1.696&0.27 &0.27&  $\pm$0.1 & $3$  \\ 
\oii\lam3727 &1.708   &0.95 &1.11& $\pm$0.3 & $8$  \\ 
\enddata
\tablecomments{
(1) Detected emission lines and their rest frame wavelength.
(2) Redshift calculated from the observed wavelength.
(3) Measured flux, relative to \hb, detector response uncorrected. 
(4) Measured flux, relative to \hb,  detector response corrected. 
(5) 1 $\sigma$ error of the line flux from Gaussian fitting.
(6) Signal-to-noise of the line flux.}
\end{deluxetable}
%%%%

At redshift 1.7, all the rest-frame optical metallicity diagnostic lines fall into the spectral coverage
 of MOIRCS zJ500 grism. We have detected all the rest frame nebular emission lines from \oii\ to \ha,  
 including the usually very weak \oiii\lam4363 line. Figure~\ref{fig:spec} shows our final two-dimensional and one-dimensional rest-frame optical spectrum of Lens22.3.   Table~\ref{tab:obs} lists the measured line fluxes relative to \hb,  the 1 $\sigma$ error, 
and S/N.  Since the metallicity studies use only line ratios, we did not  attempt to flux calibrate the spectrum, 
which is nontrivial in near-IR observations. 
 The instrument response is essentially constant for the wavelength range from 
\oiii\lam4363  to  \oiii\lam5007.  The \ha\  line lies in the region where the
response has dropped by $\sim 35\%$. The response on the expected \nii\ line  position has dropped by 
$\sim 65\%$, which may be the reason that there is no detection of \nii\  within 3$\sigma$. 
After correcting for instrumental response,  we obtain an upper limit for \nii/\ha.  
Metallicities were calculated using the following two methods.

{\it 1. Strong line diagnostics.}  In the absence of direct measurements from electron temperature,
 the ratio of strong nebular lines is used to determine the oxygen abundance. The 
well-known discrepancies between the different line ratio indicators can be removed by converting
 to the same base calibration following the methods of \citet{Kewley08}.
We use the indicator $R_{23}$\,=\,(\oii\lam3727 + \oiii\llam4959, 5007)/\hb\/
 to calculate metallicity.  In our case, the upper and lower branch degeneracy of $R_{23}$
can be broken by the upper limit of \nii/\ha\,$\,<\,$0.005.
In addition, the detection of the weak \oiii\lam4363 line almost certainly means the adoption of the lower branch.  For the actual calculation,  we use
 the lower branch calibration of \citet{McGaugh91} (M91 method) in the analytical form given by \citet{Kobulnicky99}. 
 The resulting metallicity is  \ohe\ 8.0 in the M91 method.   Converted to the \citet{KD02} method (KD02), 
 we have \ohe\ 8.1 assuming zero extinction, or 8.3 assuming a high extinction value of E(B-V)\,=\,0.4 \citep{Cardelli89,Osterbrock89}.   The high extinction value agrees with the 
 direct estimate from the \ha/\hb\ in Table~\ref{tab:obs}. However,  because  the \ha\  line lies near the edge where the response has dropped by $\sim 35\%$,  the ratio of \ha/\hb\ can be largely uncertain because it depends sensitively on the precision of correction for the detector response.  Therefore instead of using only the high extinction from \ha/\hb, we use two extreme values of extinction to quantify the range of uncertainty caused by extinction. 
 Note that the logarithm in the definition of \oh\ means that 10\% of change in the line ratio  causes only  $\sim$ 0.04 dex change in metallicity.  The uncertainty for our derived metallicity is therefore $<$ 0.1 dex prior to 
 extinction correction.

{\it 2. Direct metallicity from electron temperature $T_{e}$.}
The auroral line \oiii\lam4363 is often very weak, even in  low metallicity environments.  It is usually not observed in 
high metallicity galaxies without very sensitive, high S/N spectra \citep[e.g.,][]{Garnett04}.
The 3 $\sigma$ detection of  \oiii\lam4363 in Lens22.3  strongly suggests a low metallicity environment.
The ratio of \oiii\lam5007, 4959 and  \oiii\lam4363 
allows a direct measurement of the oxygen abundance via electron temperature $T_{e}$.   
We use the {\tt nebular} package in {\sc iraf} to derive the electron temperature
$T_e$\,=\,2.3557 $\times$ 10$^{4}$ K, assuming an electron density n$_{e}$ of 100 cm$^{-3}$ ($T_e$ is
relatively insensitive to $n_{e}$, e.g.,  for $n_{e}$\,=\,1000 cm$^{-3}$, $T_e$\,=\,2.3508 $\times$ 10$^{4}$ K).  Following the procedure of \citet{Izotov06}, we obtain \ohe 7.5 $\pm$ 0.1 assuming zero extinction; or \ohe 7.3 $\pm$ 0.1 assuming
a high extinction value of E(B-V)\,=\,0.4. 

The 0.6 dex discrepancy between the KD02 and  $T_{e}$ method is not surprising since the metallicity measured from the direct $T_e$ method is systematically lower than other methods;  such offsets can be as large as 0.7 dex \citep{Liang06, Yin07b,Nagao06}.  Regardless of the offset between the two methods, both metallicity estimates indicate that
 Lens22.3 is  low in oxygen abundance.  The evidence that Lens22.3 is a star-forming galaxy instead of a narrow line AGN comes from its position on the traditional 
 Baldwin---Phillips---Terlevich (BPT) diagram \citep{Baldwin81}.  Using the upper limit of  \nii/\ha\  and the ratio of \oiii/\hb,  Lens22.3 is  located on the leftmost region of the star-forming branch of the BPT  \nii\ diagram, excluding its possibility of being an AGN \citep{Kewley06}.

\subsection{Photometry}
We searched the {\em Hubble Space Telescope} ({\it{HST}}) archive and obtained the photometry for Lens22.3.   
The observed AB magnitudes are 23.2 $\pm$ 0.1, 23.6 $\pm$ 0.2, 23.3 $\pm$ 0.1, 23.4 $\pm$ 0.1 and 23.1 $\pm$ 0.1 in the {\textit{ACS}} {\ensuremath{g_{F475W}}}, {\ensuremath{r_{F625W}}}, {\ensuremath{i_{F775W}}}, {\ensuremath{z_{F850LP}}}, and {\textit{NICMOS}}  {\ensuremath{J_{F110W}}} bands, respectively.
After correcting for the lensing magnification of 15.5 $\pm$ 0.3 (provided by J. Richard 2009, private communication),  
the absolute  $B$ band magnitude is $M_{B}$\,=\,-18.3 $\pm$ 0.1 (Vega System). 
A simple spectral energy distribution (SED) fitting using the \verb+kcorrect v4_1_4+ code of \citet{Blanton07} yields a stellar mass of 
1.6 $\times$ 10$^{9}$ M$_{\odot}$.  Since the \verb+kcorrect+ may overestimate the stellar masses of high-redshift galaxies \citep{Blanton07},  we recalculated the stellar mass using the 
advanced version of the code 
\verb+LE PHARE+ \citep[photometric redshift and simulation package;][]{Ilbert09} based on
population synthesis models of \citet{BC03}.
The resulting stellar mass is 4.4$\pm$1.2$\times$10$^{8}$ M$_{\odot}$.

\subsection{Morphology and Kinematics}
%%%
\begin{figure}[!ht]
\epsscale{1}
\plotone{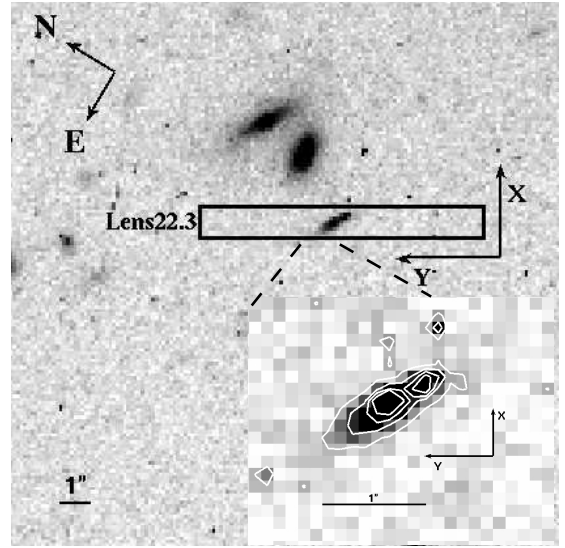}
\caption{{\it HST} rescaled F775W image for the field of Lens22.3  The image has been rescaled and registered onto the MOIRCS chip. The slit on Lens22.3 has a width of 1\,$\!\!^{\prime\prime}$ and a length of 9\,$\!\!^{\prime\prime}$.  Slit $X$ (dispersion) and $Y$ (spatial) direction are marked.  A zoomed view
of Lens22.3 with intensity contours (0.3, 0.2, 0.1, 0.04 of the peak intensity) is shown on the lower left corner.}
\label{fig:lens}
\end{figure}
%%%
%%%
\begin{figure}[!ht]
\epsscale{1}
\plotone{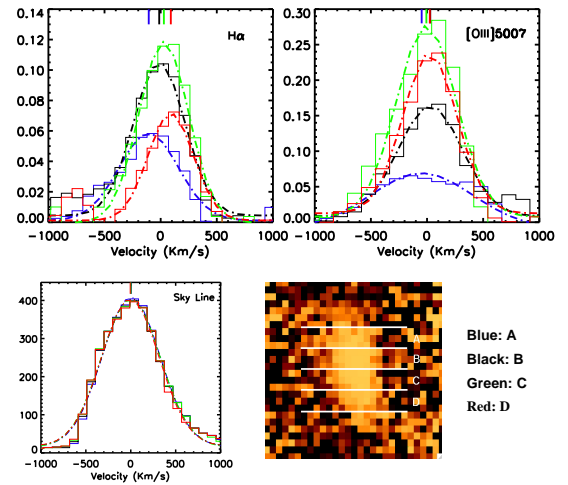}
\caption{Top two panels: velocity profile of \ha\ and \oiii\ from spectra extracted within the spatial apertures
 A, B, C, D (bottom right panel).  Bottom left panel:  velocity profile for the sky line nearest to \ha\ using the
 same apertures. Bottom right panel:   two-dimensional spectrum of \ha, spatially cut into sub-apertures A, B, C, D.
The velocity zero point is chosen to be the averaged center of the cuts.}
\label{fig:lineprof}
\end{figure}
%%%
Thanks to the excellent seeing ($\sim$  0.$\!\!^{\prime\prime}$4) 
 and the powerful magnification of gravitational lensing,  the target is spatially resolved.
 In Figure~\ref{fig:lens}, we show the {\textit{HST}} F775W image in the vicinity of Lens22.3. 
The image has been registered onto the MOIRCS chip, which has a pixel scale of 0.$\!\!^{\prime\prime}$117 pixel$^{-1}$. The MOIRCS slit layout is shown, where   
$X$ is the dispersion direction and $Y$ is the spatial direction which is also along the elongated side of the galaxy.  
On the lower right corner is a zoomed image of Lens22.3, with various intensity contours overlaid. 
It has a clear double-nuclei-like morphology.  The  nuclei are separated by $\sim$ 0.$\!\!^{\prime\prime}$4, 
corresponding to a physical scale of $\sim$3 kpc  at the redshift of $z = 1.7$. 
The intrinsic separation will be smaller than this value because the image has been elongated by lensing effects. 
 We divide the two-dimensional spectrum into subsections  A, B, C, D along the spatial direction $Y$,  using a 3 pixel bin that roughly matches the average seeing of the observation.  
We extract the spectrum for apertures A, B, C, D using the strongest line \oiii\lam5007 and \ha.  The velocity profile relative to the center of the subsections are given in Figure~\ref{fig:lineprof}.  
For comparison,  the velocity profile for the sky OH line nearest to \ha\  extracted from the same apertures is also shown. There seems to be a systematic rise in the  velocity from A to D in \ha,  but not in \oiii\lam5007. 
The misalignment of the line profile in the spatial direction and the double-nuclei like morphology may indicate that 
the system is rotating or merging.   Future follow up with integral field spectroscopy with adaptive optics is highly desirable to reveal the kinematics
of the system.
  
%%%
\begin{figure*}[!ht]
\epsscale{0.9}
\plotone{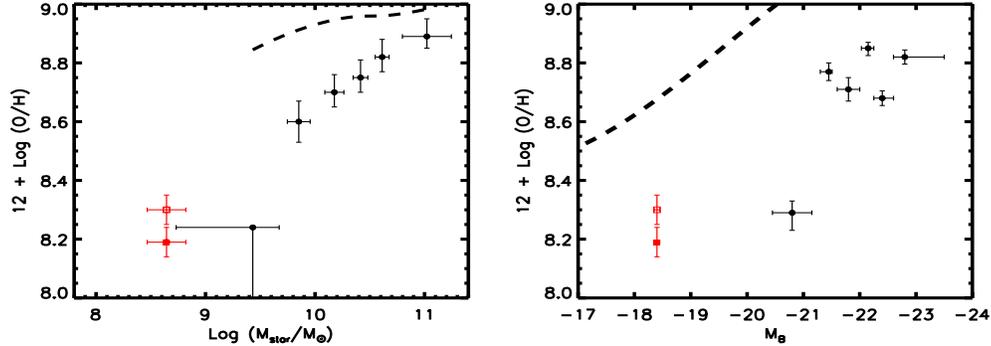}
\caption{ Left: mass---metallicity relation.   
Dash line:  local relation from the SDSS sample \citep{Tremonti04}.  
Filled black circle:  relation for galaxies z $\geq$ 2 \citep{Erb06}.   
Filled red square: Lens22.3, without extinction correction.
Empty red square: Lens22.3, after extinction correction.
Right: luminosity---metallicity relation. 
Dash line:  local relation from SDSS sample \citep{Tremonti04}.  
Filled black circle:  relation for galaxies z $\geq$ 2 \citep{Erb06}.   
Filled red square: Lens22.3, without extinction correction.
Empty red square: Lens22.3, after extinction correction.
All metallicities have been recalculated to the same base of KD02 \citep{Kewley08}.
}
\label{fig:ml}
\end{figure*}
%%%

\subsection{Mass-Metallicity Relation}
In Figure~\ref{fig:ml},  we show the mass/luminosity---metallicity relation for the local \citep{Tremonti04} and high-redshift 
\citep{Erb06}  star-forming galaxies.  All metallicities in Figure~\ref{fig:ml} have been rescaled to the same KD02 method
using the calibration of \citet{Kewley08}. 
The local sample of \citet{Tremonti04} is based on $\sim$ 53,000 Sloan Digital Sky Survey (SDSS) star-forming galaxies and 
the \citet{Erb06} sample is based on composite spectra of 87 rest-frame UV-selected star-forming galaxies at z $\geq$ 2.   Lens22.3 is consistent with the mass/luminosity---metallicity relation of the high z sample.  We evaluate the effect of extinction
 assuming a high extinction value of E(B-V)\,=\,0.4.
   Extinction causes  metallicity to rise by $\sim$ 0.1 dex,  but
 it does not significantly change the position of Lens22.3 on the mass---metallicity diagram.  
Lens22.3 is currently the lowest mass system with direct measurement on the mass---metallicity relation diagram;  Lens22.3 extends the luminosity---metallicity relation to the fainter end by more than an order of magnitude, demonstrating the capability of using gravitationally lensed samples
to probe intrinsically fainter and less massive systems than are found in existing samples.

\section{Summary}
We reported the discovery of a low metallicity star-forming galaxy at $z = 1.7$.  
We detect the temperature sensitive auroral line \oiii\lam4363
which allows us to derive  metallicity using the direct $T_{e}$ method for the first time at high redshift.
The low metallicity and low luminosity/mass of the $z = 1.7$ star-forming 
galaxy place it at the lowest end of the mass/luminosity---metallicity relation, extending the 
$z>2$ luminosity-metallicity relation by more than an order of magnitude. 
Our on-going near-IR spectroscopic survey of lensing clusters aims to obtain 
metallicities for more objects like Lens22.3
 in the redshift range 1.5 $<$ z $<$ 3.

%%%%%%%%%%%%%%%%%%
\acknowledgments 
We want to thank the referee for her/his valuable comments and suggestions that significantly
improved this Letter.  We thank Johan Richard for providing us with the lensing magnification of Lens22.3, and Youichi Ohyama for his help with MOIRCS data reduction.  We also thank Katelyn Allers for useful discussion on Telluric correction methods and C.-J. Ma for running the \verb+LE PHARE+  code for us.   We recognize and acknowledge the very significant cultural role and reverence that the summit of Mauna Kea has always had within the indigenous Hawaiian community.  We appreciate the opportunity to conduct observations from this sacred mountain.

{\it Facilities:} {\it Subaru} (MOIRCS)

\end{document}